\documentclass[a4paper,12pt]{article}

\usepackage[english]{babel}
\usepackage[utf8x]{inputenc}
\usepackage[T1]{fontenc}
\usepackage{authblk}
\usepackage[dvipdfmx]{graphicx}
\usepackage{natbib}
\usepackage{here}
\usepackage{setspace}
\usepackage{multirow}
\usepackage{lmodern}
\usepackage{amsmath, amssymb}
\usepackage{bm}
\usepackage{pdflscape}
\usepackage{authblk}
\usepackage{color}
\usepackage[table]{xcolor}

\usepackage{soul}

\usepackage{booktabs}
\usepackage{afterpage}
\usepackage{textcomp}
\usepackage{float}
\usepackage{enumerate}
\usepackage{rotating}

\definecolor{babyblue}{rgb}{0.54, 0.81, 0.94}
\definecolor{babypink}{rgb}{0.96, 0.76, 0.76}

\usepackage[margin=1in]{geometry}

\usepackage{amsmath}
\usepackage{graphicx}
\usepackage[colorlinks=true, allcolors=blue]{hyperref}

\usepackage{caption}
\usepackage{subcaption}

\usepackage{lineno}

\begin{document}

    \begin{center}
        \vspace*{1cm}
        \large
        \textbf{Integrating automated acoustic vocalization data and point count surveys for estimation of bird abundance}\\
         \normalsize
           \vspace{5mm}
         Jeffrey W. Doser\textsuperscript{1, 2}, Andrew O. Finley\textsuperscript{1, 2, 3}, Aaron S. Weed\textsuperscript{4}, 
         Elise F. Zipkin\textsuperscript{2, 5}
         \vspace{5mm}
    \end{center}
    \small
         \textsuperscript{1}Department of Forestry, Michigan State University, East Lansing, MI, 48824, USA \\
         \textsuperscript{2}Ecology, Evolution, and Behavior Program, Michigan State University, East Lansing, MI 48824, USA \\
         \textsuperscript{3}Department of Geography, Environment, and Spatial Sciences, Michigan State University, East Lansing, MI, 48824, USA \\
         \textsuperscript{4}Northeast Temperate Inventory and Monitoring Network, National Park Service, Woodstock, VT 05091, USA \\
         \textsuperscript{5}Department of Integrative Biology, Michigan State University, East Lansing, MI 48824, USA \\
          \noindent \textbf{Corresponding Author}: Jeffrey W. Doser, email: doserjef@msu.edu; ORCID ID: 0000-0002-8950-9895 \\
          \noindent \textbf{Running Title}: Integrating acoustic and point count data

\section*{Abstract}

\begin{enumerate}
    \item Monitoring wildlife abundance across space and time is an essential task to study their population dynamics and inform effective management. Acoustic recording units are a promising technology for efficiently monitoring bird populations and communities. While current acoustic data models provide information on the presence/absence of individual species, new approaches are needed to monitor population abundance, ideally across large spatio-temporal regions. 
    \item We present an integrated modeling framework that combines high-quality but temporally sparse bird point count survey data with acoustic recordings. Our models account for imperfect detection in both data types and false positive errors in the acoustic data. Using simulations, we compare the accuracy and precision of abundance estimates using differing amounts of acoustic vocalizations obtained from a clustering algorithm, point count data, and a subset of manually validated acoustic vocalizations. We also use our modeling framework in a case study to estimate abundance of the Eastern Wood-Pewee (\textit{Contopus virens}) in Vermont, U.S.A.
    \item The simulation study reveals that combining acoustic and point count data via an integrated model improves accuracy and precision of abundance estimates compared with models informed by either acoustic or point count data alone. Improved estimates are obtained across a wide-range of scenarios, with the largest gains occurring when detection probability for the point count data is low. Combining acoustic data with only a small number of point count surveys yields estimates of abundance without the need for validating any of the identified vocalizations from the acoustic data. Within our case study, the integrated models provided moderate support for a decline of the Eastern Wood-Pewee in this region. 
    \item Our integrated modeling approach combines dense acoustic data with few point count surveys to deliver reliable estimates of species abundance without the need for manual identification of acoustic vocalizations or a prohibitively expensive large number of repeated point count surveys. Our proposed approach offers an efficient monitoring alternative for large spatio-temporal regions when point count data are difficult to obtain or when monitoring is focused on rare species with low detection probability. 
\end{enumerate}

\noindent \textbf{Keywords}: acoustic recording units, data fusion, joint likelihood, avian, imperfect detection, false positives, N-Mixture model
\newpage 

\section*{Introduction}

Monitoring species abundance across large spatio-temporal regions is a crucial task for informing wildlife management (e.g., \citealt{Farr2019}) and assessing the influence of forest management, climate change, and other anthropogenic stressors on population dynamics \citep{Furnas2019}. The need for understanding species abundance patterns has led to widespread development of monitoring designs and associated modeling techniques that address imperfect detection of individuals during wildlife sampling, such as distance sampling \citep{buckland1993} and repeated counts \citep{Royle2004}. These methods have been widely successful for estimating wildlife population dynamics and informing management across relatively small spatial scales \citep{Zipkin2014}. However, cost and time required to implement such methods limits their use for understanding drivers of species' abundances across large spatio-temporal regions. Thus, improved monitoring designs and associated statistical tools that can reliably estimate wildlife population dynamics across large spatio-temporal regions have the potential to provide more detailed information for informed wildlife management and conservation \citep{Miller2019}. 

Acoustic recording units (ARUs) have recently emerged as a viable option for monitoring acoustically-communicating wildlife \citep{Gibb2019}. Recent advances in ecoacoustics \citep{Farina2018} have improved our understanding of how anthropogenic and biotic sounds interact across the environment (e.g., \citealt{Doser2020}) and how the soundscape can be a rich source of ecological monitoring data \citep{Sueur2014}. A variety of indices that summarize acoustic space are used to monitor biodiversity \citep{Sueur2014} and quantify disturbance effects (e.g., \citealt{Burivalova2018, Doser2020EI}). 

The usefulness of acoustic indices as ecological indicators is limited by a lack of species-specific information associated with recorded sounds \citep{Gibb2019}. This has led to an interest in extracting species-specific information from acoustic recordings and using such information to estimate wildlife distributions, especially for birds \citep{shonfield2017autonomous}. Most often, detection/non-detection data are extracted from acoustic recordings and used in occupancy models \citep{MacKenzie2002} to infer species distributions \citep{Furnas2015}. A second approach uses fully automated machine learning algorithms to determine whether or not a species is present in a specific recording (e.g., \citealt{Ducrettet2020}). These automated methods show promise, but currently are not reliable enough to inform species specific information without identification of recorded calls to account for false positives (i.e., recording a species as present when it is truly absent). 

Bird density can be estimated from a dense array of ARUs using spatial capture-recapture \citep{Efford2009} as well as distance sampling \citep{Darras2018a, yip2019sound}. However, establishing dense arrays of ARUs is expensive, and distance sampling requires calibration experiments that might not be feasible for large-scale surveys. Alternatively, there is widespread interest in integrating acoustic data with point count survey data to improve bird monitoring programs \citep{shonfield2017autonomous}, particularly in boreal North America \citep{pankratz2017potential, van2017paired}. These approaches integrate point count data with ARU data using distance sampling and removal sampling to estimate correction factors that account for differences in the effective detection radius of point counts and ARUs \citep{solymos2013calibrating, van2017paired}. While such approaches provide detection-corrected estimates of density, they require intensive analysis of acoustic recordings and information on the detection distance and time of detection. In addition, they rely on the assumption that detection distances are observed with no error \citep{yip2017experimentally}, which is rarely the case when sampling birds \citep{alldredge2007field}. However, more recent approaches that integrate multiple sources of data together in a likelihood framework can lead to higher precision and accuracy of abundance estimates compared with analyses of either data set separately \citep{Miller2019}. These approaches assume a set of shared parameters across data likelihoods, and thus each data set contributes to estimation of the shared parameters while each likelihood also contains specific parameters to account for the detection processes unique to each data set. Thus, alternative data integration methods for bird abundance estimation that are not reliant on distance sampling assumptions could provide an attractive alternative when distance sampling data are unavailable or unreliable. 

We propose two integrated models leveraging both ARU and point count data to deliver accurate and precise estimates of abundance. Our first model integrates acoustic vocalizations obtained from a clustering algorithm, point count data, and a subset of manually verified detections from the clustering algorithm as in \cite{Chambert2018}. Our second model uses only the acoustic vocalizations and point count data and does not require validation of any acoustic data. Our integrated models consist of two sub-models for the acoustic and point count data linked through a joint likelihood framework \citep{Miller2019}. The acoustic model is an extension of the model of \cite{Chambert2018} to estimate abundance rather than occupancy by using acoustic vocalizations derived from a clustering algorithm and accounting for false positives directly in the model. The second component is the standard N-mixture model \citep{Royle2004} used to estimate abundance from repeated counts. We perform a simulation study to compare the ability of the acoustic model, the N-mixture model, and the two integrated models to estimate abundance under a wide range of realistic scenarios.  We subsequently apply all four models to estimate abundance and trends of the Eastern Wood-Pewee (\textit{Contopus virens}) in a National Historical Park in Vermont, U.S.A. from 2013-2015. Recent evidence suggests declines in Eastern Wood-Pewee across its range \citep{vtBirds}, which has led to widespread interest in monitoring the species and understanding the drivers of the decline \citep{audobonVermont}. The integrated models provide reliable estimates of abundance with minimal manual validation of acoustic data under a wide range of scenarios, suggesting our approach is an efficient monitoring alternative when point count data are sparse or when monitoring is focused on rare, elusive species.

\section*{Modeling Framework}

We compare the performance of four models: 

\begin{itemize}
    \item Model AV: uses acoustic (A) vocalizations obtained from a clustering algorithm and a subset of manually validated (V) vocalizations.
    \item Model C: uses point count (C) data.
    \item Model AC: uses acoustic vocalizations obtained from a clustering algorithm and point count data.
    \item Model AVC: uses acoustic vocalizations obtained from a clustering algorithm, a subset of manually validated vocalizations, and point count data.
\end{itemize}

We describe Model AV and Model C separately then detail their connection via data integration to form Model AC and Model AVC. While we could distinguish a Model A that only uses acoustic vocalizations obtained from a clustering algorithm, this model would not provide any mechanism to discriminate between false positive and true positive detections, which is known to lead to biased estimates when estimating occupancy and/or abundance (Chapter 7; \citealt{keryRoyle2020}), and so we do not consider Model A further.

\subsection*{Model AV}

Suppose an ARU was set up at $i = 1, \dots, R$ sites (i.e., defined as the area in which the ARU can detect vocalizations) and a total of $j = 1, \dots, J$ acoustic recording surveys were obtained at each of the $R$ sites. We suppose the $R \times J$ recordings are analyzed using a semi-automated clustering algorithm for acoustic vocalizations such as Kaleidoscope Pro \citep{wildlifeAcoustics}. Alternative clustering or machine learning algorithms could also be used. Kaleidoscope Pro takes acoustic recordings as inputs and determines a set of candidate vocalizations based on the timing and signal energy of the syllables in the vocalization. Coefficients from a Discrete Cosine Transformation on each candidate vocalization are extracted into a feature vector used to define clusters of similar vocalizations. Vocalizations are assigned to individual clusters using a Hidden Markov Model that is initiated with a k-means clustering algorithm. Each cluster is subsequently identified to the species level by a human observer, and only the clusters corresponding to the species of interest are used in subsequent modeling \citep{wildlifeAcoustics}. This semi-automated process yields the number of detected vocalizations, $v_{i, j}$, at each site $i$ during survey $j$. $v_{i, j}$ provides partial information on the true latent abundance of individuals at each site $i$, denoted as $N_i$. We assume abundance arises from a Poisson process according to 

\begin{equation}\label{abundance}
    N_i \sim \text{Poisson}(\lambda_i),
\end{equation}

where $\lambda_i$ is the expected abundance of individuals at site $i$. We model $\lambda_i$ according to

\begin{equation}\label{abundance-covs}
  \text{log}(\lambda_i) = \beta_0 + \bm{X}_{1, i} \cdot \bm{\beta}_1, 
\end{equation}

where $\beta_0$ is an intercept, $\bm{X}_{1, i}$ is a row-vector of spatially-varying covariates and $\bm{\beta}_1$ is a vector of regression coefficients. Alternative distributions (e.g., negative binomial) can be used to model $N_i$ if the Poisson distribution is inadequate to describe abundance (e.g., because of overdispersion).  

Because the clustering algorithm is imperfect and not all individuals present vocalize during sampling, $v_{i, j}$ can consist of both false positives and false negatives. False positives arise from the clustering algorithm incorrectly assigning an acoustic detection to a cluster that represents vocalizations from the species of interest. False negatives may arise from the clustering algorithm failing to detect a vocalization present in the recording, or from individuals not producing any vocalizations during the survey $j$. To accommodate the preponderance of false negatives and false positives, we model $v_{i, j}$ using a Poisson hurdle model \citep{dorazio2013estimating}. The hurdle model is a mixture of a Bernoulli process and a zero-truncated Poisson process, where the Bernoulli process models the probability of detecting at least one vocalization at a given site $i$ during survey $j$, and the zero-truncated Poisson process models the intensity of vocalizations at site $i$ during survey $j$ conditional on at least one vocalization being detected. More specifically, let $y_{i, j} = 1$, when $v_{i, j} > 0$, and 0 otherwise. To model the probability of detecting at least one vocalization at site $i$ and survey $j$, we have 

\begin{equation}\label{bernoulli-A}
    y_{i, j} \sim \text{Bernoulli}(\pi_{i, j}),
\end{equation}

where $\pi_{i, j}$ is the probability of detecting at least one vocalization at site $i$ during survey $j$. We model $\pi_{i, j}$ as 

\begin{equation}\label{detection-A}
    \text{logit}(\pi_{i, j}) = \alpha_0 + \alpha_1 \cdot N_i + \bm{X}_{2, i, j} \cdot \bm{\alpha_2}, 
\end{equation}

where $\alpha_0$ represents the probability of detecting a vocalization at an unoccupied site on the logit scale, $\alpha_0 + \alpha_1$ represents the probability (on the logit scale) of detecting at least one vocalization at a site with a single individual, and $\bm{\alpha}_2$ is a vector of regression coefficients describing the effect of survey-level covariates, $\bm{X}_{2, i, j}$, on detection. To complete the hurdle model, for all $i$ and $j$ where $y_{i, j} = 1$, we have 

\begin{equation}\label{ztPoisson-A}
    v_{i, j} \sim \text{ztPoisson}((\delta_{i, j} \cdot N_i + \omega) \cdot y_{i, j}),
\end{equation}

where ztPoisson is a zero-truncated Poisson distribution, $\omega$ is the average number of false positive detections, and $\delta_{i, j}$ is the average rate of detections per-individual. Thus, on average, each survey $j$ at site $i$ where at least one vocalization was detected (i.e., $y_{i, j} = 1$) will consist of $\delta_{i, j} \cdot N_i + \omega$ vocalizations, of which $\delta_{i, j} \cdot N_i$ are true vocalizations and $\omega$ are false-positives. $\delta_{i, j}$, the per-individual true positive rate, is a compound process determined by the vocalization rate of individual birds and by the ability of the clustering algorithm to detect vocalizations \citep{Chambert2018}. Bird vocalization rates are known to vary across both space and time \citep{digby2014temporal, ord2017does}, and thus we model $\delta_{i, j}$ according to

\begin{equation}\label{delta-A}
    \text{log}(\delta_{i, j}) = \gamma_0 + \bm{X}_{3, i, j} \cdot \bm{\gamma}_1,
\end{equation}

where $\gamma_0$ is the intercept and $\bm{\gamma}_1$ is a vector of regression coefficients describing the effect of survey-level covariates, $\bm{X}_{3, i, j}$, on the true positive detection rate.

As currently constructed, the model requires an additional data source to separately estimate the false positives from true positives. Accordingly, a small subset of the detected vocalizations (e.g., 10\%) are manually verified by listening to the identified vocalizations in Kaleidoscope Pro Software and determining if the vocalization is a true positive or false positive. This validation process yields two additional data sources: $n_{i, j}$, the number of manually checked vocalizations at site $i$ and survey $j$, and $k_{i, j}$, the number of manually checked vocalizations at site $i$ and survey $j$ verified to be true. We include the auxiliary information in our model using the hypergeometric distribution \citep{Chambert2018, keryRoyle2020} where we have 

\begin{equation}\label{auxiliary-A}
\begin{split}
    K_{i, j} & \sim \text{Binomial}(\text{tp}_{i, j}, v_{i, j}) \\
    k_{i, j} &\sim \text{Hypergeometric}(K_{i, j}, Q_{i, j}, n_{i, j}),
\end{split}
\end{equation}

where $\text{tp}_{i, j}$ is the true-positive rate of detections at site $i$, defined as $\frac{\delta_{i, j} \cdot N_i}{(\delta_{i, j} \cdot N_i) + \omega}$, $K_{i, j}$ is the total number of true detections at site $i$ and survey $j$, and $Q_{i, j}$ is the number of false positive detections in the clustering algorithm (i.e., $v_{i, j} - K_{i, j}$) at site $i$ and survey $j$. 

\subsection*{Model C}

Suppose $l = 1, \dots, L$ point count surveys were performed at $i = 1, \dots, I$ sites (i.e., the area in which birds are detected during the survey). Assuming population closure across the $L$ surveys, the standard model for estimating abundance from repeated counts is the N-mixture model \citep{Royle2004}, defined as 

\begin{equation} \label{nmix}
   c_{i, l} \sim \text{Binomial}(N_i, p_{i, l}), \\
\end{equation}

where $N_i$ (which follows Equation \ref{abundance}) is the latent abundance at site $i$, $c_{i, l}$ is the number of individuals detected at site $i$ and survey $l$, and $p_{i, l}$ is the probability of detecting an individual at site $i$ and survey $l$. Detection of individuals, $p_{i, l}$, can vary across space and survey replicates according to 

\begin{equation} \label{nmix-detection}
   \text{logit}(p_{i, l}) = \phi_0 + \bm{X}_{4, i, l} \cdot \bm{\phi}_1,  
\end{equation}

where $\phi_0$ is the intercept and $\bm{\phi}_1$ is a vector of regression coefficients describing the effect of survey-level covariates $\bm{X}_{4, i, l}$ on detection.

\subsection*{Integrated Models AC and AVC}

Model AV and Model C both provide information on expected population abundance, $\lambda_i$, and subsequently the true latent abundance, $N_i$, at each site $i$. Site-level abundance $N_i$ in both models can be interpreted as ``superpopulation'' abundance (e.g., the number of birds whose home ranges overlap the sampling units) as the detection probability components both depend on an individual being: (1) present at site $i$ during the survey, (2) available for detection during the survey $j$ or $l$, and (3) detected by the algorithm/observer \citep{nichols2009}. By integrating the two data sets we can use the information from both the acoustic data and point count data to improve  estimation of $N_i$ and the underlying parameters ($\beta_0$ and $\bm{\beta_1}$) that govern expected abundance, $\lambda_i$. ARUs and point counts could be implemented at the same sites in the study region at the same time, or the two survey approaches could be implemented at different locations and times within the breeding season as long as site-level population closure is reasonable. To integrate the acoustic and point count data, we use a joint likelihood framework to account for the different detection processes that govern the acoustic and point count data. We assume the two detection models are independent, both conditional on the true latent abundance $N_i$ at each site $i$ \citep{Miller2019}. This is equivalent to assuming the probability of detecting an individual bird during a point count survey is independent of the probability the acoustic recorder detects a vocalization \citep{keryRoyle2020}. This assumption enables the joint likelihood to be written as the product of the individual likelihoods. More specifically, the likelihood for Model AVC is the product of the Poisson hurdle model for the acoustic data (Equations \ref{bernoulli-A} and \ref{ztPoisson-A}), the hypergeometric process for the validated subset of the acoustic data (Equation \ref{auxiliary-A}), and the binomial process for the count data (Equation \ref{nmix}). The likelihood for Model AC is the product of the Poisson hurdle model (Equations \ref{bernoulli-A} and \ref{ztPoisson-A}) and the binomial count process (Equation \ref{nmix}). Model AC does not incorporate the manually validated data ($k_{i, j}$ and $n_{i, j}$) and instead relies on the information in count data to separate true positives from false positives in the acoustic data. The joint likelihoods for both models are conditional on a single, shared latent abundance process (Equation \ref{abundance}), which enables the sharing of information between the two data types.

\section*{Model Implementation}

We estimated the parameters in all models with a Bayesian framework using Markov chain Monte Carlo (MCMC) \citep{hooten} and vague priors for all parameters (Supplemental Information S1).  MCMC samplers were written in JAGS \citep{Plummer03jags} within the \texttt{R} statistical environment \citep{r} using the \verb+jagsUI+ package \citep{jagsUI}. Model convergence was assessed using the Gelman-Rubin R-hat diagnostic, with convergence presumed when R < 1.1 \citep{brooks1998}. 

\section*{Simulations}

We performed a simulation study to assess the ability of the four models to estimate abundance. We varied parameters under a wide range of ecologically realistic values to provide a broad assessment of model performance. We simulated covariates that affect abundance ($\lambda_i$), true positive rate ($\delta_{i, j}$), and detection processes ($\pi_{i, j}$, $p_{i, k}$) as standard normally distributed variables. 

We generated data with $J = 10$ acoustic recordings at each site. The total number of sites was set to either 50 or 100. We varied the ratios between the number of acoustic sites, $R$, and the number of point count sites, $I$, resulting in three scenarios: (1) $R = I$; (2) $R = \frac{I}{2}$; (3) $\frac{R}{2} = I$. We set $L = \{3, 5\}$ (low and high replication), $\beta_0 = \{0.2, 1\}$ (rare and abundant species), $\alpha_1 = \{1.2, 3\}$ (low and high probability of detecting at least one acoustic vocalization), resulting in 48 different simulation scenarios. We simulated the effect ($\beta_1$, $\alpha_2$, $\gamma_1$, $\phi_1$) of a single covariate ($X_{1, i}, X_{2, i, j}, X_{3, i, j}, X_{4, i, l}$) on $\lambda_i$, $\pi_{i, j}$, $\delta_{i, j}$, and $p_{i, l}$, respectively. The following parameters were used to generate 100 data sets for all 48 simulation scenarios: $\beta_1 = 0.3$, $\alpha_0 = -2.19$, $\alpha_2 = 0.2$, $\omega = 3$, $\gamma_0 = 1.0$, $\gamma_1 = 0.5$, $\phi_0 = \alpha_0 + \alpha_1$, and $\phi_1 = 0.2$. We did not vary these parameters because exploratory simulations revealed similar model performance across different values of these parameters. For the models AV and AVC, we simulated manual verification of 20\% of the vocalizations. Model performance was assessed by comparing the ability of the models to recover the abundance intercept, $\beta_0$, and covariate effect, $\beta_1$. We used the median relative bias to assess the level of bias in the parameter estimates, and used the width of the $95\%$ credible interval (CI) to assess the precision around those estimates. 

We further assessed the performance of ARU data to estimate abundance by determining how abundance estimates changed as the number of point count sites increased. We generated 100 data sets with the following parameters: $R = I = 50$, $J = 10$, $L = 4$, $\beta_0$ = 2, $\beta_1 = 0.3$, $\alpha_0$ = -2.19, $\alpha_1 = 3$, $\alpha_2 = 0.2$, $\gamma_1 = 1$, $\gamma_2 = 0.5$, $\phi_0 = 0.81$, $\phi_1 = 0.2$, and $\omega = 3$. These parameter values corresponded to an easily detectable, abundant species. For each of the 100 generated data sets, we varied the number of randomly selected point count sites between 5, 10, 20, 30, and 50. The model was run for the five sampling scenarios for each data set, resulting in a total of 500 simulations. We used these simulations to assess the performance of Model AC, which allowed us to determine the number of point count sites needed to separate the false positive detections from the true detections in the acoustic data without manual validation. For all simulations and each model, we ran three chains of 10,000 iterations with a burn-in period of 3000 iterations, an adaptive period of 5000 iterations, and a thinning rate of 2, resulting in 10,500 samples from the posterior distribution from which we summarized the model results. 

\section*{Case Study}

\begin{figure}
    \centering
    \includegraphics[trim=0cm 2.25cm 0cm 2.25cm, clip, width = 13cm]{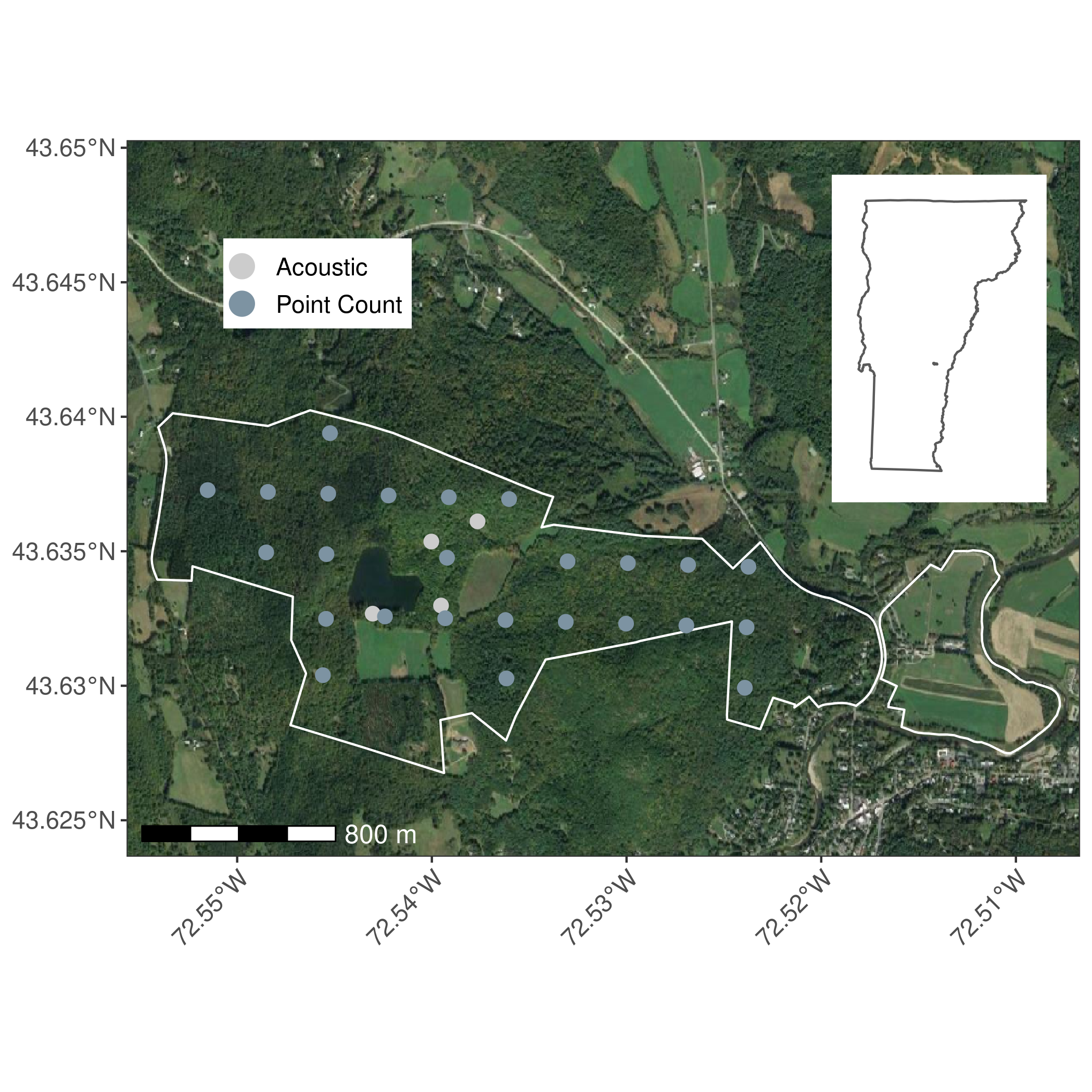}
    \caption{Locations of acoustic recorders and point counts in Marsh Billings Rockefeller (MABI) National Historical Park (bounded region) in Vermont, U.S.A. Inset map shows the location of MABI (black point) in Vermont.}
    \label{fig:mabi}
\end{figure}

We assessed the performance of the four models to estimate annual abundance and trends in Eastern Wood-Pewee abundance from acoustic and point count data collected in the Marsh-Billings-Rockefeller (MABI) National Historical Park in Vermont, U.S.A. (Figure~\ref{fig:mabi}). The Eastern Wood-Pewee is a migratory bird species found throughout deciduous forests in the eastern US. Recent evidence suggests declines in the species across its range \citep{vtBirds, audobonVermont}. During the peak breeding periods from 2006-2019 (May-June), volunteers performed point count surveys annually at 25 sites separated by 200-250m throughout the park where they recorded all individual birds seen or heard within 10 minutes (full details on the survey sites and protocols can be found in \cite{netnBird2015}). 

We sought to incorporate additional acoustic data through our proposed models to improve the accuracy and precision of Eastern Wood-Pewee abundance estimates. ARUs were deployed within the park from 2013-2015 and thus we focus our analyses on this time period. SongMeter SM2 acoustic recorders \citep{wildlifeAcoustics} were placed at four sites in MABI and were scheduled to record for ten minutes at the beginning of each hour continuously from March to November of 2013-2015, regardless of weather. These ARUs were originally deployed for different monitoring purposes, which resulted in the four ARUs being placed near a pond. For estimating bird abundance, we only used recordings from June 1 - June 14 during the hours of 5:00AM-8:00AM. This resulted in four, 10-minute recordings across 14 days at each of the four sites for each of the three years (or 672 total recordings). Due to problems with the recording devices and memory constraints, we were able to use 443 (or 66\%) 10-minute recordings. We used Kaleidoscope Pro software \citep{wildlifeAcoustics} to identify clusters of similar vocalizations contained in the recordings, and subsequently listened to recordings from each cluster and only used vocalizations from the clusters identified as EAWP vocalizations. This process produced the number of vocalizations, $v_{i, j}$, identified as Eastern Wood-Pewee by the clustering algorithm in survey $j$ at site $i$. We manually validated 30\% of the identified EAWP vocalizations for use in models AV and AVC. 

We estimated EAWP abundance using each of the four models previously described. However, there was overdispersion in the acoustic vocalization data that could not be accounted for by the Poisson hurdle model. As such, we extended the model to a negative binomial hurdle model, where the zero-truncated negative binomial process was specified as a mixture of a zero-truncated Poisson and gamma distribution. More specifically, we have 

\begin{align}\label{neg-binom}
    v_{i, j} &\sim \text{ztPoisson}((\delta_{i, j} \cdot N_i + \omega) \cdot y_{i, j} \cdot \epsilon_{i, j}) \\
    \epsilon_{i, j} &\sim \text{Gamma}(a_{v}, a_{v}),
\end{align}

where $\epsilon_{i, j}$ is extra variability in the number of vocalizations detected at site $i$ during survey $j$, and $a_v$ is an estimated parameter controlling the amount of overdispersion. We modeled $\delta_{i, j}$ according to $\text{log}(\delta_{i, j}) = \gamma_{0, i, \text{DAY}_j}$, where $\gamma_{0, i, \text{DAY}_j}$ is a normally distributed random effect (e.g., $\gamma_{0, i, \text{DAY}_j} \sim \text{Normal}(0, \tau_{\text{DAY}})$) to account for variation in the vocalization production rate across days. The probability of detecting at least one vocalization ($\pi_{i, j}$) was modeled as $\text{logit}(\pi_{i, j}) = \alpha_0 + \alpha_1 \cdot N_i$, while we assumed that the detection probability for the count data was constant ($p_{i, l} = p$) after exploratory data analysis revealed little variation in repeated counts across sites. We modeled expected abundance as a linear function of year in order to assess the ability of the four models to estimate trends across the three years. More specifically, abundance is modeled as $\text{log}(\lambda_{i, t}) = \beta_0 + \beta_1 \cdot \text{YEAR}_{t}$, where $\lambda_{i, t}$ is the abundance at site $i$ during year $t$ and YEAR$_{t}$ is the specific year corresponding to $t$. We standardize YEAR$_t$ to have mean 0 and standard deviation of one to facilitate comparison with previous estimated trends in EAWP abundance. This case study presents a unique test of the models as the acoustic data are temporally dense ($J = 42$ surveys) but spatially sparse (only $R = 4$ sites), while the count data are temporally sparse (only $L = 2$ surveys at some sites) but spatially dense ($I = 25$ sites). We ran three chains of 200,000 iterations with a burn in period of 60,000 and a thinning rate of 50, resulting in 8,400 samples from the posterior distribution of the parameters for each model. Model fit was assessed using a Bayesian p-value approach \citep{Gelman1996} with the Freeman-Tukey statistic as a measure of fit. A separate Bayesian p-value was computed for each data set used in a given model.

\section*{Results}

\subsection*{Simulations}

\begin{figure}
    \centering
    \includegraphics[width = 15cm]{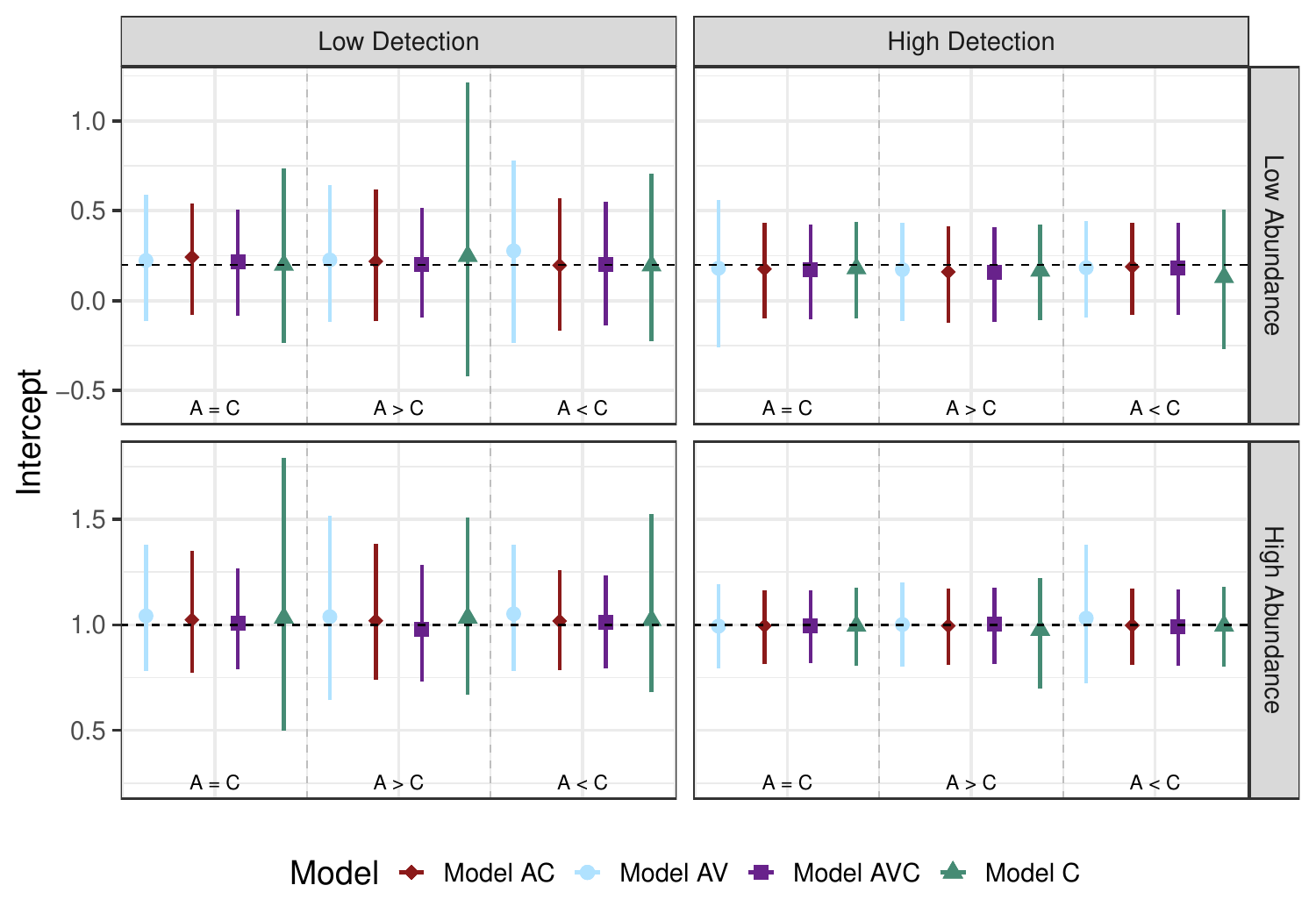}
    \caption{Simulation results on the accuracy and precision of abundance intercept estimates ($\beta_0$) among models with varying detection probability and expected abundance. Points show the posterior median and lines represent the 95\% credible interval for 100 simulated data sets. All data sets were simulated using $J = 10$ and $L = 5$. A = C corresponds to $R = I = 50$, A > C corresponds to $R = 50, I = 25$, and A < C corresponds to $R = 25, I = 50$.}
    \label{fig:covSimulations-beta0}
\end{figure}

\begin{figure}[!ht]
    \centering
    \includegraphics[width = 13cm]{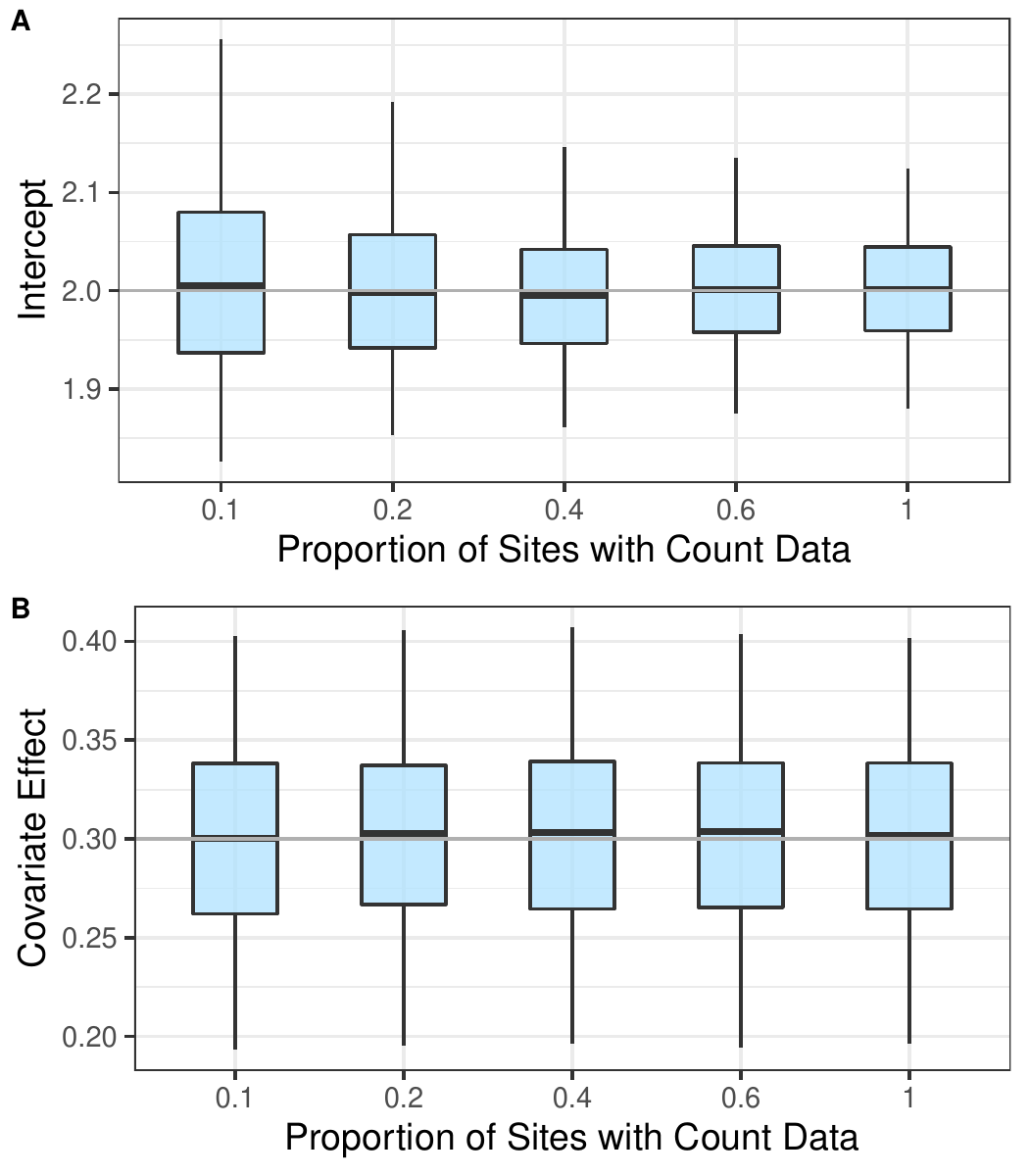}
    \caption{Estimated abundance intercept ($\beta_0$) and covariate ($\beta_1$) with 50 ARU sites and a variable number of point count sites using Model AC. The boxplots represent the median across 100 simulations of the median (dark line), 50 \% CI (box), and 95\% CI (whiskers).}
    \label{fig:randomSamplingCov}
\end{figure}

Across all 48 simulated scenarios, the integrated models (AC and AVC) were consistently more accurate, more precise, and converged more often than the single data source models (Figure ~\ref{fig:covSimulations-beta0}). The integrated models were more consistently precise than Model AV. Model C provided similar estimates to both integrated models when detection was high, but the integrated models showed substantial improvements in precision over Model C when detection was low. The median relative bias of the abundance intercept ($\beta_0$) across all simulation scenarios was 1.18\% for Model AV, -1.09\% for Model C, 0.56\% for Model AC, and 0.52\% for Model AVC. The median 95\% CI width of $\beta_0$ was 0.53 for Model AV, 0.63 for Model C, 0.42 for Model AC, and 0.40 for Model AVC. Together, this suggests only minor improvements in accuracy and precision in Model AVC compared to Model AC. All models were able to accurately and precisely estimate the covariate effect, $\beta_1$, with minor improvements gained from the two integrated models (Supplemental Information S1: Figure S2). Further, Model AC yielded unbiased estimates of abundance even with small numbers of high quality repeated point counts (Figure \ref{fig:randomSamplingCov}, Supplemental Information S1: Table S1), suggesting abundance can be estimated using ARU data combined with small amounts of point count data without any validation of acoustic vocalizations from the clustering algorithm. 

Some simulated data sets with a small number of observations did not allow for separation of the detection process from the abundance process and subsequently led to a lack of convergence. Model AV also failed to converge for some simulated data sets where abundance was high and most sites were occupied, suggesting acoustic data alone may not yield reliable estimates of abundance for abundant species. Across all simulations, 90\%, 81\%, 94\% and 97\% of models converged, respectively, for models AV, C, AC, and AVC. Not only did the integrated models improve parameter estimates, they also increased the rate of convergence in cases where each individual data set alone did not provide enough information to separately estimate the detection and abundance processes.

\subsection*{Case Study}

All four models predicted a decreasing trend in EAWP abundance across the three years, although support for the decline varied across models. Model C estimated $\beta_0$ with low precision as a result of the small number of repeated surveys within a year \citep{kery2015applied}, while Model AV estimated $\beta_1$ with low precision as a result of the small number of sites. Integrating both data types in Models AC and AVC resulted in substantial improvement in precision of both $\beta_0$ and $\beta_1$, with Model AVC being the most precise. Bayesian p-values indicated successful model fit for all models, with all Bayesian p-values having means ranging between 0.335 and 0.491 (See Supplemental Information S1: Table S2). The improved performance of Model AVC over Model AC in this case study differs from the only slight improvements in precision found in the simulations, suggesting manual validation of some acoustic vocalizations might be necessary if the point count data are of low quality (i.e., small number of repeat surveys). 

\begin{figure}
    \centering
    \includegraphics[width = 15cm]{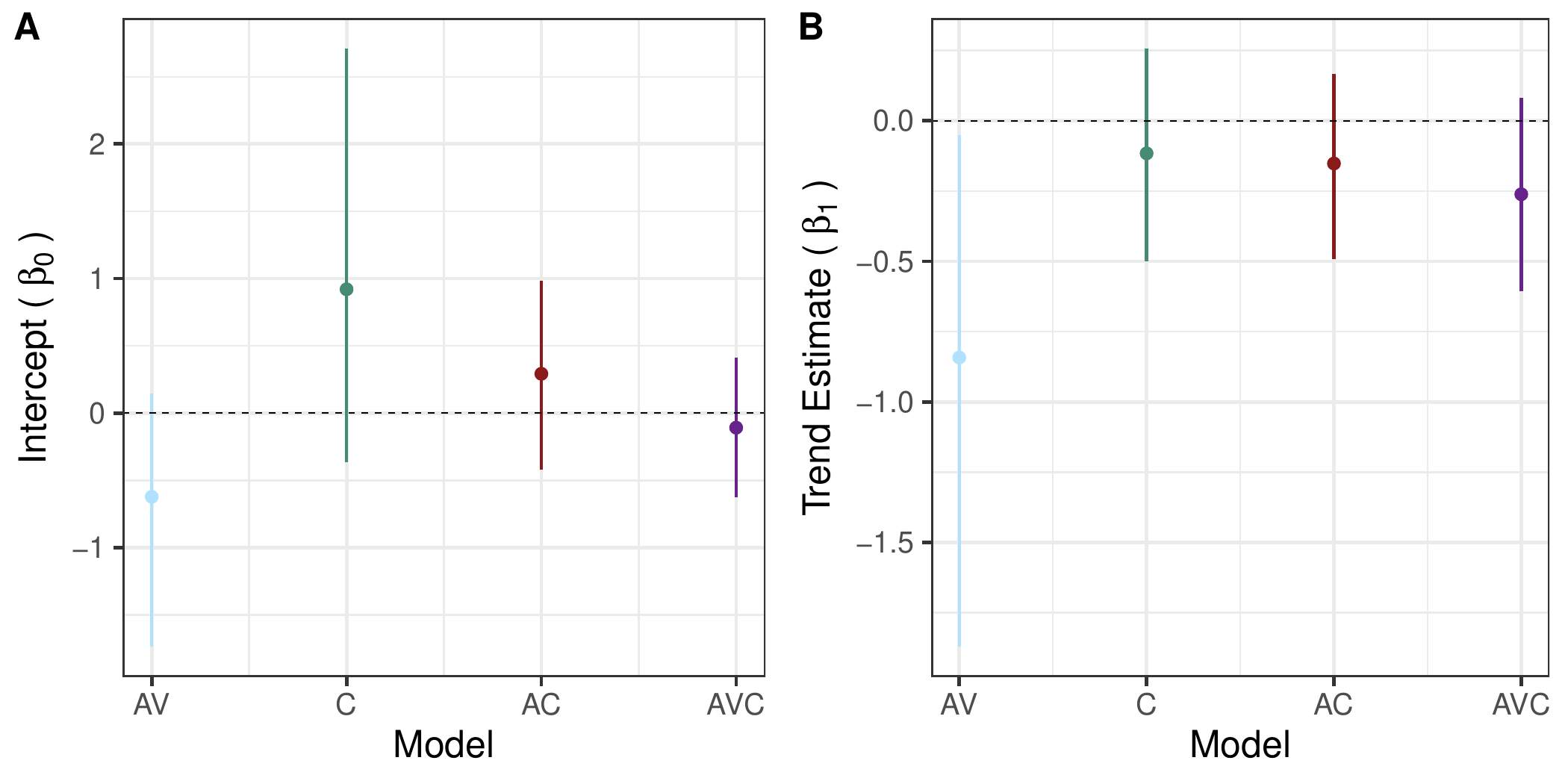}
    \caption{Comparison of model performance for estimating the abundance intercept (A) and yearly trend (B) of the Eastern Wood-Pewee. Points represent the posterior median and lines represent the 95\% credible interval.}
    \label{fig:eawp-abundance}
\end{figure}

\section*{Discussion}

Combining acoustic recording and point count data in an integrated modeling framework consistently led to accurate and precise estimates of abundance across 48 simulation scenarios. Our simulation study incorporated a wide-range of realistic scenarios varying in the ease of target species detection, abundance, and number of acoustic and point count sites. Combining the two data types within an integrated model yielded improved parameter accuracy and precision over analysis of a single data source. These results are consistent with several recent studies that demonstrate the advantage of leveraging multiple data types for estimating species distributions \citep{Miller2019}.

Despite the general improvement in accuracy, precision, and convergence when integrating acoustic recording and point count data, there are situations when reasonable estimates can be obtained using a single data source. For example, the median 95\% CI of the abundance intercept ($\beta_0$) for Model AV across all 48 simulation scenarios always contained the true parameter value, suggesting abundance can be estimated reasonably well with only acoustic data (provided a subset of the data are manually verified). However, when abundance is large (e.g., three birds per site) the model may fail to converge when most sites are occupied, as there is less information to separate true positive detections from false positives. Interestingly, Model AV provided nearly identical estimates for the covariate effect on abundance compared to the integrated models (Supplemental Information S1: Figure S2). Thus, we consider Model AV a reasonable method for estimating abundance when abundance is low or when inference focuses on assessing covariate effects on abundance. 

Abundance estimates from Model C, the N-mixture model, are known to improve as the detection probability, number of sites, and number of repeated surveys at each site increases \citep{Royle2004, kery2015applied}. If such high quality point count data exist, combining count data with acoustic data will yield little improvements in accuracy and precision of abundance estimates, as evidenced by the close correspondence between Model C and the two integrated models when detection probability was high in our 48 simulations (Figure~\ref{fig:covSimulations-beta0}, Supplemental Information S1: Figure S2). However, our results also suggest that using a combination of acoustic recorders and point count surveys can provide an alternative to increasing the number of point count sites to achieve a specified level of precision in abundance estimates. Integrated approaches to monitoring could potentially be more cost-effective and feasible given the specific monitoring objectives or logistical constraints \citep{shonfield2017autonomous}. 

Across all simulations, Model AVC yielded only minor improvements in precision compared to Model AC, which does not require any manual verification of acoustic data. We further showed that when combining acoustic recordings with only a small number of repeated point count surveys, Model AC can provide accurate and precise estimates of abundance (Figure~\ref{fig:randomSamplingCov}). However, the Eastern Wood-Pewee case study showed Model AVC had substantial improvements in precision compared to Model AC, which suggests manual validation of acoustic vocalizations may be necessary to achieve desired precision levels if the point count surveys are limited in temporal replication. These results together suggest the following approach to estimating bird abundance: (1) implement a large number of ARUs across the spatial region; (2) perform a small number of high-quality (i.e., multiple repeat surveys) point count surveys at a subset of those sites; (3) use a clustering algorithm to obtain vocalizations from the ARU recordings; and (4) use Model AC to estimate abundance and any desired covariate effects. 

Neither of the two data sources include information on the exact spatial location of individual birds, and thus our approach does not enable the estimation of an effective detection radius (e.g., \citealt{van2017paired}) for the two data types. Thus, an implicit assumption of our integrated modeling approach is that the maximum detection radius of the acoustic data is approximately equal to that of the point count data (i.e., an ARU site covers roughly the same area as a point count survey). If the maximum detection radius of the ARU is smaller than the detection radius of the point count data, an offset can be included in the detection model of the acoustic data to integrate the two data types. Further, our models allow the detection probability of both the ARU and point count data sets to vary across both space and time as shown in our simulation study and case study. By using separate likelihoods for the acoustic and point count data, our models effectively account for differences in detection probability across the two data types.

The Eastern Wood-Pewee case study shows the integrated models provide more precise estimates of the abundance intercept ($\beta_0$) and trend parameter ($\beta_1$) than those produced through analysis of either the acoustic data or the point count data alone (Figure~\ref{fig:eawp-abundance}). The acoustic data are only available at four sites, which leads to large uncertainty in the trend estimate. While the count data have better spatial replication, they lack temporal replication at sites within a given year, leading to high uncertainty in $\beta_0$. The two integrated models use the temporal replication of the acoustic data to improve the estimation of $\beta_0$ and the spatial replication of the count data to provide more precise estimates of the yearly trend ($\beta_1$), revealing the benefit of incorporating both acoustic and point count data when each data type alone cannot yield robust estimation of abundance. Because the ARUs were not implemented as part of the original study design and were somewhat clustered near wet areas within the park, this could potentially lead to conclusions from the acoustic data that are not indicative of the entire park. However, all models showed at least moderate support for a decreasing trend in EAWP abundance from 2013-2015, which aligns with the negative temporal trend from 2006-2019 of EAWP detected at MABI from a removal sampling effort \citep{doser2020trends}. 

The ability of N-mixture models to yield unbiased estimates of true latent abundance ($N_i$) is highly dependent on model assumptions and proper model specification \citep{link2018robustness, barker2018reliability, knape2018sensitivity, duarte2018fitting}. Such assumptions include the binomial assumption for repeated counts, unmodeled variation in detection probability, closure of the population, and unmodeled variation in abundance \citep{link2018robustness}. Thus, we suggest using measures of goodness of fit (GoF), such as Bayesian p-values, to assess the reliability of the specific forms of the abundance and detection models. If such assumptions are a concern, the integrated models can still provide reliable information on covariate effects influencing relative abundance across space \citep{barker2018reliability}. Our approach is also dependent on the assumption that the manually validated vocalizations are validated without error. Potential concerns regarding the accuracy of these validations could be alleviated by having multiple observers validate the recordings. Despite these assumptions, the integrated models do not rely on perfect estimation of detection distances (which is often not achieved when using distance sampling for estimation of bird density \citep{alldredge2007field}), providing an alternative approach to other methods integrating ARU and point count data \citep{van2017paired, pankratz2017potential}. 

Improved estimation of wildlife abundance is critical to the study of their population dynamics and subsequent conservation. We demonstrated that combining data from acoustic recordings and point count surveys in an integrated model can provide accurate and precise estimates of wildlife abundance, even without the need to validate acoustic vocalizations in certain conditions. Given the relative ease with which acoustic recorders can be deployed across a large variety of habitats and regions, acoustic recordings, in combination with a small number of high-quality point count surveys, can serve as a reliable method for estimating abundance of acoustically-communicating species. 

\section*{Authors' Contributions}

JWD developed the models, analyzed the data, and led writing of the manuscript. AOF and EFZ provided critical input on the methodology. ASW assisted in data management and the case study. All authors contributed critically to the drafts and gave final approval for publication. 

\section*{Acknowledgements}

We thank Terri Donovan and her team (University of Vermont) and  Brian Mitchell (NPS) for deploying and coordinating the ARU data collection. We declare no conflicts of interest. This work was supported by: National Science Foundation grants DMS-1916395, EF-1253225, and DBI-1954406; U.S. National Park Service.

\section*{Data Availability}

All data and code associated with this manuscript are available at \url{https://doi.org/10.5281/zenodo.4532174} \citep{dataAvail}.

\bibliographystyle{apalike}

\end{document}